# Distributed Gesture Controlled Systems for Human-Machine Interface

Hans Johnson and Jafar Saniie

*Embedded Computing and Signal Processing (ECASP) Research Laboratory (http://ecasp.ece.iit.edu)*
*Department of Electrical and Computer Engineering*
*Illinois Institute of Technology, Chicago IL, U.S.A.*

*Abstract* — This paper presents the design flow of an IoT human–machine touchless interface. The device uses embedded computing in conjunction with the Leap Motion Controller to provide an accurate and intuitive touchless interface. Its main function is to augment current touchscreen devices in public spaces through a combination of computer vision technology, event-driven programming, and machine learning. Especially following the COVID-19 pandemic, this technology is important for hygiene and sanitation purposes for public devices such as airport, food, and ATM kiosks where hundreds or even thousands of people may touch these devices in a single day. A prototype of the touchless interface was designed with a Leap Motion Controller housed on a Windows PC exchanging information with a Raspberry Pi microcontroller via internet connection.

## I. INTRODUCTION

Touchless interfaces have been a sought-after goal in human-machine interaction, but it stands as a fact that modern implementations of the technology have performed too poorly to see a widespread integration of such technologies. However, the Leap Motion Controller offers a ray of hope into this field as an incredibly small but powerful hand recognition module. It has a relatively simplistic hardware design that is complimented by very powerful tracking software. This hardware/software codesign allows for incredible classification speeds in real-time, making the Leap Motion Controller a true flagship device for hand recognition. The data itself is also accessible through the Leap Motion's open-source API, meaning integration of AI based approaches are very possible for the device. Combined with innovative approaches, the Leap Motion Controller is a gateway device to changing the way we interact with machines and information.

Versus other similar systems, the Leap Motion Controller offers a unique form factor module designed specifically for tracking hands. Opposed to the existing solutions, the Leap Motion Controller is much more accurate and is better suited for PC applications. The human hand is capable of attaining an accuracy around 0.4 mm on average, and the Leap Motion Controller was determined to have an overall average accuracy of 0.7 mm per axis [1]. The closest competitor, the Kinect, sees a static accuracy of about 1.5 cm [2]. Compared to other controllers in this price range, the Leap Motion Controller is unparalleled as a consumer product.

Additionally, when subject to Fitt's law – a predictive model of human movement for human-computer interaction and ergonomics based on moving a pointer to a target [3] – it was found that the controller had an error rate of 7.2% versus 2.8% with a standard mouse device [4]. This means that inaccuracies happened around three times as much than a standard mouse, though this study was completed in 2014 and does not account for Leap Motion's subsequent software updates and did not employ any stabilization practices in coding [4]. It is also worth noting the accuracy is subjective to the application, therefore these metrics of accuracy were approached with caution.

An overview of the proposed prototype system is shown in Figure 1. The actual distributed system human machine interface (HMI) is a combination of the Leap Motion Controller, a Raspberry Pi, a camera, and a display utilizing IoT and wired connections. The camera is optional because it serves as an added safety feature for applications such as ATMs and security systems. The Leap Motion Controller receives user inputs and sends the information to the local server (PC) as events. The server then communicates control commands directly to the client (Raspberry Pi) over the network, and the client is responsible for the UI the user interacts with. The proposed system would have multiple distributed HMIs connected to a single server which could eventually be replaced with a cloud; however, this paper introduces a prototype that serves as a proof of concept.

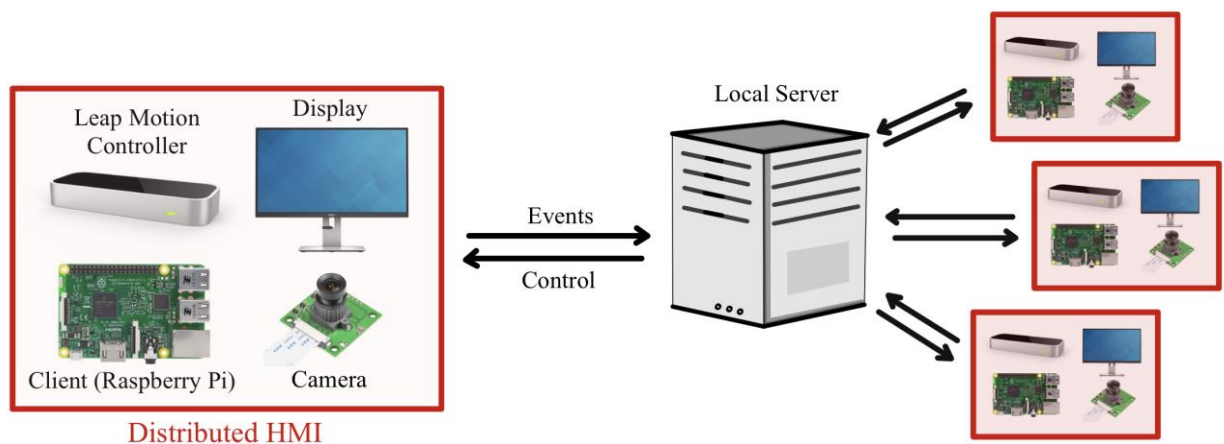

Figure 1. Design Flow for Distributed Gesture Controlled HMI System

A distributive-based approach could lead to projects integrating this software into touchless UI for public kiosks, medical applications, education, and vehicles – especially with the rising popularity of larger dash screens such as those exhibited in luxury vehicles like Tesla, BMW, and Mercedes. The innovative potential of the device still stands to make way for new types of applications geared toward touchless interfacing with machines. It is especially relevant for the sake of hygiene in public spaces with publicly used kiosks such as menus in fast food restaurants, ATMs, and ticket kiosks where hundreds or even thousands of people touch these machines in a single day.

## II. SYSTEM DESIGN

The prototype was designed to assume control of a computer cursor using exclusively gesture control. The system was partitioned into two areas: server side and client side. The Leap Motion Controller was wired directly to the PC acting as a server. A Raspberry Pi microcontroller was deployed on the client side with bidirectional communication to the PC via WebSocket. It has been previously shown that multiple Leap Motion Controllers can run on one PC/Server [5]. This shows the ability to utilize computationally intense devices on one server deployed on a cheaper client-side device, though only one controller was used for this prototype. Likewise, multiple client-side devices may be connected to one server as well.

### A. Server Software Design

The general software design for the server-side device was designed to interpret information directly from the Leap Motion Controller and send commands to the client device. The server-side software was created with the goal of replacing touch controls with a gesture-controlled cursor. In order to define the exclusive state of the hand interacting with the Leap Motion device, the algorithm is comprised of the following state variables:

- **Detect Index Finger**: The index finger is identified for assuming control of the cursor. Tracking points are assigned to the fingertip and joints and the operation is maintained until the finger leaves the bounding box.
- **Click**: Represented with a screen tap in the air by the index finger, the click functions the same as a left click on a mouse.
- **Hold**: The controller tracks the tips of all other fingers and performs a left click hold.
- **Release**: Invoked by the *Hold* state variable, the left click is released when all fingers are brought back into the hand.
- **Scroll Up**: A counterclockwise circular motion of the index finger triggers the window to scroll up.
- **Scroll Down**: A clockwise circular motion of the index finger triggers the window to scroll down.

The state variables are invoked in the server, and the updates are sent to the client over the network in real time. This process is explained in the flowchart below in Figure 2. Around 200-300 frames of data are interpreted every second and only the coordinates and commands are sent to the client. This data is also siphoned through error correction methods to ignore very minute movements of the hand to eliminate errors from unstable hands or accidental movement when performing gestures like the *Click* gesture.

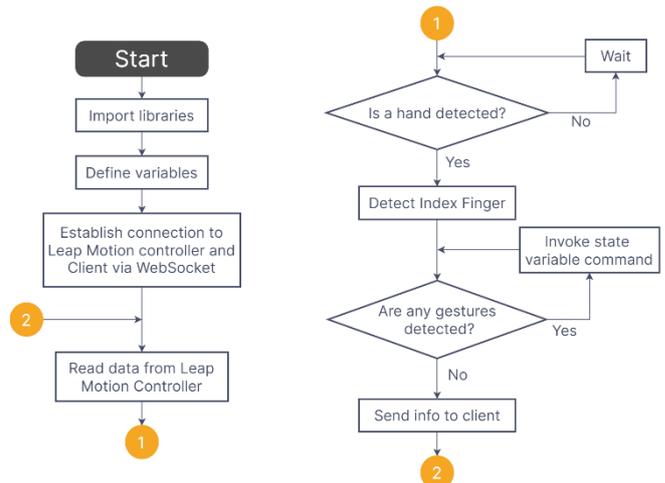

Figure 2. Flowchart of Touchless Interface Server

### B. Client Software Design

On the client side, the software interprets commands directly from the server via WebSocket and interacts with the user via the display/user interface (UI). Once connected to the server, the client awaits gesture commands to be sent from the server through the input stream. The information sent in the input stream is sent in coordinates that the cursor can take in and output on the display, which is a massive reduction of information and interpretation that is done between the Leap Motion Controller and the client. The five commands for Click, Hold, Release, Scroll Up, and Scroll Down are assigned to commands that the cursor must perform. An IO Exception is put into place in case the client has trouble communicating with the server.

Even though hand data is interpreted separate from the client, the flow of operation is still determined by events driven by user action. A user is interacting and responding to the information that shows up on the screen rather than commanding the device. The operation of the client-side software is shown in Figure 3 below.

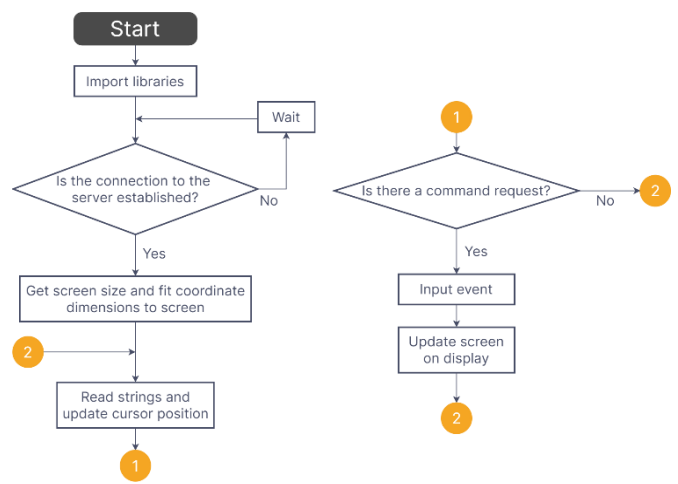

Figure 3. Flowchart of Touchless Interface Client

In terms of security, no security protocols were implemented other than establishing a connection via WebSocket to the IP of the Raspberry Pi. The data was not encrypted, which could prove trouble on a machine such as an ATM with very sensitive information. Further developments are needed for encryption, but for the sake of the prototype it was not implemented.

## C. Hardware Design

The most important piece for the hardware design was the Leap Motion Controller. It is a state-of-the-art real-time hand tracking USB peripheral device that covers a hemispherical area of about one meter. It was shipped by UltraLeap in July of 2013 and has since seen active development and updates to its software by the Leap Motion community. The device itself is extremely small, fast, and accurate with an average cost of about $80 with its SDK. It can be used for applications with Windows, Linux, and MacOS, though its libraries can only be compiled for 32-bit and 64-bit systems with a processor equivalent to Intel's i3/i5/i7 CPUs. The controller requires a high-powered processor because it creates and analyzes 200-300 frames of data per second and cannot be run on something like an entry-level ARM processor built for a microcomputer. Its primary purpose was for productivity for gesture-controlled computer applications and augmenting HMIs, integration into enterprise-grade hardware solutions/displays, and attachment to virtual reality headsets for VR/AR/XR research and development as well as prototyping [6].

The hardware for the device is rather simplistic and minimalist. Contained in the module there are two PCB boards, two wide-angle IR cameras, three illuminating IR LEDs, and a Programmable System-on-Chip (PSoC) among other auxiliary components. The device is powered by a USB 3.0 micro-B 2/3 hybrid data cable which also serves as the data line to send inputs to the local server.

The controller is capable of tracking hands within a 3D interaction zone created with the dual IR camera's stereoscopic image rendering. The interaction zone is 2 ft x 2 ft x 2 ft (8 ft3) normal to the module as seen in Figure 4 and can discern 27 distinct hand elements (including bones and joints) and track them even when obscured by other parts of the hand [6].

The beauty of the device lies within the software. The data is sent from the controller to the PC for all the processing. The frames of data are then processed in real-time frame-by-frame to generate tracking information based on the raw sensor data. This is one reason why the PC requires a i3/i5/i7 processor or equivalent. Once the tracking information is generated, it is sent via transport protocol to the appropriate API that manages frame history and provides functions and classes in an Object Orientated Programming (OOP) structure/library [7].

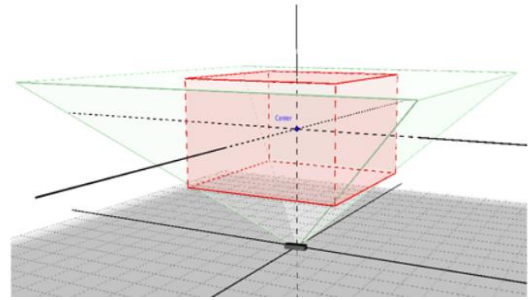

Figure 4. Designated Leap Motion interaction box [6]

Outside of the controller, a laptop PC was utilized as the server in the prototype. In the future, we would like to research options of replacing the PC with a cloud server. The Raspberry Pi 4 B was utilized as the client because many of the kiosks in public spaces utilize similar (if not the exact model) microcontrollers for their operations.

The project itself was constrained to provide real-time feedback, meaning that it had to maintain at least 30 fps to prove the efficacy of this project. This was dependent on two things: the data interpretation and network speed. The data interpretation worked because the interpretation of the frames was done on the PC with an adequate processor, and the coordinates (not the frames themselves) from these interpreted hand gestures were sent to the Raspberry Pi. The other commands that were utilized that were not exclusively gesture interpreted were easily handled by the Raspberry Pi as they are on the order of user input. The project was done over a sluggish network speed of 15 Mbps download and around 10 Mbps upload speeds, but still had virtually no latency from user input. A large delay would come from a network dropout, though the data would still be processed by the server meaning this could be circumvented with a Bluetooth connection.

The design to utilize the WebSocket for communication drastically reduces the complications of network requests. This also makes it efficient to have an implementation completed in real time. A flowchart for the overall hardware architecture is given in Figure 5.

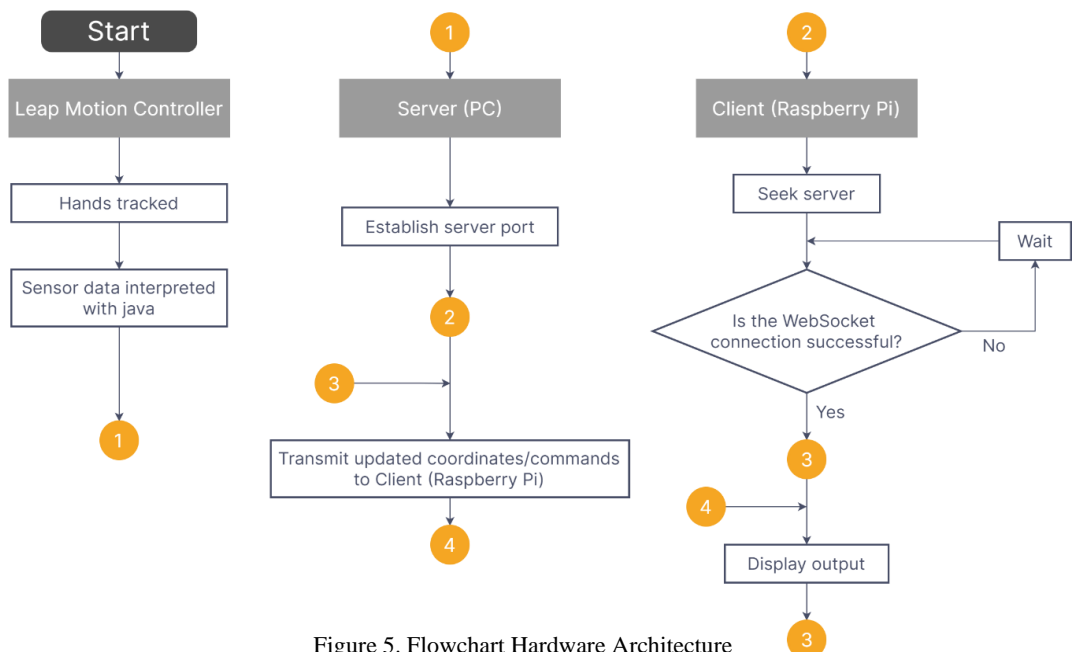

Figure 5. Flowchart Hardware Architecture

## III. IMPLEMENTATION

The implementation section discusses the development of the prototype device. It is solely based on the Leap Motion Controller, a laptop PC acting as a server, a Raspberry Pi 4 B acting as a client, and other peripheral devices for the client. The camera was not implemented in this prototype.

### A. Hardware Implementation

Considering the hardware constraints of the project, the hardware design shown in Figure 6 is proposed. The Leap Motion Controller is plugged into the USB peripheral port of the laptop computer with a 3 ft USB micro-B 2/3 hybrid data cable which delivers 5V DC power. The controller draws about 320 mA, but the dynamically driven LEDs may reduce the current draw to around 200 mA when the LEDs are dimmed depending on a hand's distance from the sensor. This prevents images from becoming saturated and keeps the image quality high.

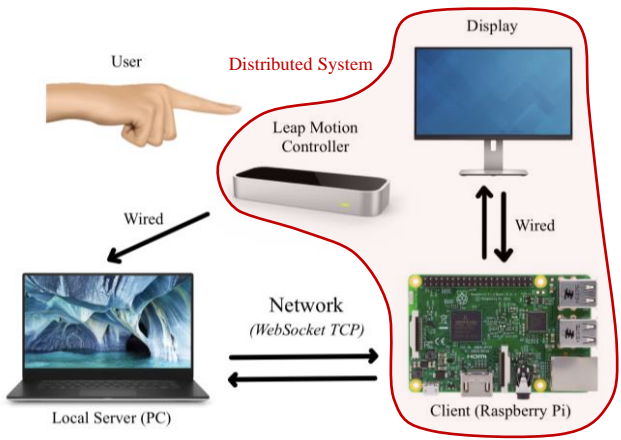

Figure 6. Overview of Touchless Interface Prototype

The Leap Motion Controller features two monochromatic IR cameras tracking at 850nm (infrared = 780nm to 1mm wavelength) and 3 LEDs that flash continuously at very high speeds. The cameras themselves are 640 x 240-pixels, or 0.15 MP each spaced 40mm apart, typically operate at 120 Hz, and can capture an image within 1/2000th of a second.

The information was interpreted and sent via WebSocket to a Raspberry Pi 4 B microcontroller that was connected to all the peripheral devices a user would interact with. The Raspberry Pi 4 is a $55 microcontroller with a Quad core Cortex-A72 ARM v8 64-bit SoC processor that can run at 1.5GHz. It comes in 2GB, 4GB, or 8GB DDR4 SDRAM models, and has 2 USB 3.0 and 2 USB 2.0 ports. It has 2 micro-HDMI ports, 2-lane MIPI DISI display and CSI camera ports, ethernet capability, 2.4GHz and 5.0 GHz IEEE 802.11ac wireless internet, Bluetooth 5.0, BLE, and a microSD slot for loading the OS and data storage. It is powered by a 5V min 3A power supply via USB-C. The 8GB model was used for this prototype.

### B. Server and Client Software Implementation

Regarding the state variables considered for the software implementation, the software was built to maximize the efficiency of both data interpretation and communication to mitigate the overhead across the server and client devices. The two codes located on the server and client respectively were built in the Java programming language, taking advantage of the portability and object-orientated nature.

The options available for the API were JavaScript, Unity, C#, C++, Java, Python, Objective-C, and Unreal. Java 7 was utilized and run by Eclipse on the PC. The Leap Motion v3 SDK was used due to compatibility, as later versions of the SDK only allowed for C and C++ API integration. The Gestures Library from the UltraLeap developers was utilized for the commands for gesture control. This library has 39 methods for interpreting information from the hands in the controller, including methods for hands, fingers, frames, gestures, and more.

At first, Python was considered to be a good option for processing information. However, as Python is not a compiled language, and the results were determined to be generally faster and more efficient in Java. Java is very Object Oriented based, meaning when working with a web socket it was easy to invoke classes on the client from data interpreted by the server. Other options that may have been faster or would provide more versatility were C and C++, but these methods seemed too complex to implement for the time frame of the project.

Between the client and server, a WebSocket connection was created versus other wireless connections due its bidirectional, full-duplex communication capabilities. It can facilitate message passing over a single TCP/IP socket connection, giving it the abilities of a User Datagram Protocol (UDP) with the consistency of a Transmission Control Protocol (TCP). Hypertext Transfer Protocol (HTTP) is used as the first transport mechanism while the TCP is kept alive after the HTTP response is received. This enables low-latency real-time transmission between the server and client, allowing for optimal results with event-driven programming in mind.

## IV. RESULTS

There were two goals for the proof of concept: program gestures and create a cursor from tracked index finger. The gestures were based off the vectors created by the Leap motion SDK, and the ones utilized was the normal vector from the palm, directional vectors, and vectors from the joints of the fingers. From the gestures library, only the *ScreenTapGesture* and *CircleGesture* were used for this project to control the mouse click and the mouse scroll. The click and hold function were done by looking at the fingers and checking if the pinky and ring finger were extended.

The validation for the prototype was demonstrated through functionality of the cursor. Figure 7 shows the setup of the proposed prototype with the cursor highlighted on the display and the diagnostic visualizer on the local server. For the other state variables (click, hold, release, scroll up, scroll down), navigating a webpage was implemented to test their functions. The internet browser on the Raspberry Pi was opened with the click function, and the scroll function was utilized by scrolling up and down the browser's homepage. Additionally, the hold and release function were demonstrated by holding the top of the opened webpage, dragging the window to a different position on the screen, then releasing the window.

As long as the network connection was maintained, all the functionalities were successful and done in real time with minimal latency. If the connection was broken, the cursor would stop in the place of the last coordinates received. When the connection was reestablished, the cursor would jump to the newest coordinates transmitted. The latency of the device was tested under conditions where the network bandwidth was at 10%.

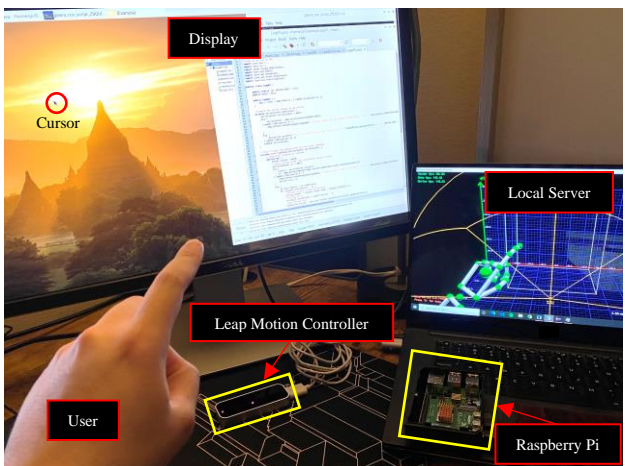

Figure 7. Touchless Interface Prototype Setup

For the system to work, the server and client had to run their own respective programs. The Leap Motion API was also running on the local server, though its data was being directly transmitted to the server through a wired connection. The data was taken directly from the Leap Motion API and then translated into the Java programming language and was interpreted through the Eclipse IDE as seen in Figure 8.

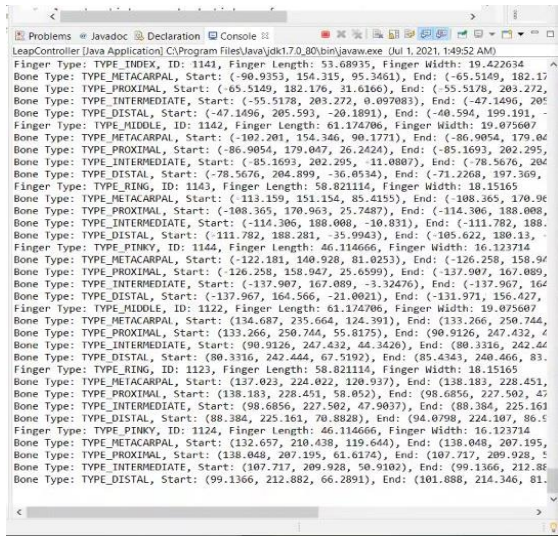

Figure 8. Java Interpretation of Leap Motion Data using Eclipse IDE

The necessary extracted frame data was sent as coordinates to the client which interacted directly with the display. This was happening in real-time with a latency of less than 20ms. There were a few errors in accuracy, as previously discussed according to Fitt's Law there was a predicted error of three times that of a standard mouse. Though the error seemed to be on par or slightly better than that prediction, refining an application for specific purposes can minimize that error with having limited functionality.

To increase the accuracy of the system, smoothing and stabilization techniques were applied to the detected index fingertip. The position of the cursor would lag behind the actual fingertip position based on the measured velocity of the fingertip. It was calibrated such that very small movements did not cause the cursor to move, such as natural shaking in a user's finger when holding their hand in front of the screen. This allowed for more accurate screen-taps and placement of the cursor on the screen.

## V. DISCUSSION AND CURRENT RESEARCH

It was demonstrated that although the non-AI integrated prototype was successful, gesture control needs to be seamless for it to be accepted as a mainstream technology. Though the accuracy of the system is subjective to the application, the most notable increase in accuracy would be seen in the positioning of the cursor and screen-tap functions. Currently, a Deep Neural Network (DNN) model is being explored to observe what different screen taps look like. We have also found multiple Leap Motion Controllers can connect to one server, though a limit has not been identified.

Currently, there exist many databases for the Leap Motion Controller completed in research. In one study, a database was formed from 100 participants completing 12 gestures, and a 3D recognition model was built on a Convolutional Neural Network (CNN) to recognize 2D projections of 3D space [8]. Another study describes a several methods for "air writing" classification for English characters using the 6D motion gesture (6DMG) dataset with Long Short-Term Memory (LSTM) and CNN algorithms to achieve an accuracy of 99.32% [9]. Extensive American sign language data sets also exist built on the Leap Motion Controller [10].

## VI. CONCLUSIONS

In this paper, a fully functional prototype of a touchless interface was designed and realized using a gesture-controlled HMI with IoT capabilities. This device is meant to be deployed as a distributed system that can be easily paired with embedded computing systems that are widespread in public/commercial use. The software and hardware designs were successfully implemented using only four modules: the Leap Motion Controller, a local server (PC), a client (Raspberry Pi), and a display. Java programming language was utilized to bring the system to fruition.